\documentclass[12pt]{article}

\newcommand{\Mdm}{\mbox{Mdm2}}
\newcommand{\pft}{\mbox{p53}}
\newcommand{\pftp}{\mbox{p53-P}}
\newcommand{\siah}{\mbox{Siah}}
\newcommand{\arf}{\mbox{Arf}}
\newcommand{\atm}{\mbox{Atm}}
\newcommand{\atmp}{\mbox{Atm-P}}
\newcommand{\dam}{\mbox{$S_D$}}
\newcommand{\nbs}{\mbox{Nbs1}}
\newcommand{\nbsp}{\mbox{Nbs1-P}}
\newcommand{\Mdmarf}{\mbox{Mdm2-Arf}}
\newcommand{\Mdmpft}{\mbox{Mdm2-p53}}
\newcommand{\Mdmpftp}{\mbox{Mdm2-p53-P}}
\usepackage{graphicx}
\usepackage{setspace}
\begin{document}
\pagestyle{plain}
\bibliographystyle{plain}

\title{ A \pft~Oscillator Model of DNA Break Repair Control}

\author{Vijay Chickarmane, Ali Nadim,  Animesh Ray and Herbert M. Sauro~\footnote{Corresponding Author :Herbert M Sauro, Keck Graduate Institute,
535 Watson Drive, Claremont, CA 91711, Phone: (909) 607 0377, Fax:
(909) 607 8086, E-mail: hsauro@kgi.edu}\\Keck Graduate Institute,
535 Watson Dr, Claremont, CA 91711}

\maketitle

\begin{abstract}
\noindent The transcription factor \pft~is an important regulator of cell
fate. Mutations in \pft~gene are associated with many cancers. In response
to signals such as DNA damage, \pft~controls the transcription of a series
of genes that cause cell cycle arrest during which DNA damage is repaired,
or triggers programmed cell death that eliminates possibly cancerous cells
wherein DNA damage might have remained unrepaired. Previous experiments
showed oscillations in \pft~level in response to DNA damage, but the
mechanism of oscillation remained unclear. Here we examine a model where
the concentrations of \pft~isoforms are regulated by \Mdm2, \arf, \siah,
and $\beta\mbox{-catenin}$. The extent of DNA damage is signalled through
the switch-like activity of a DNA damage sensor, the DNA-dependent protein
kinase \atm. This switch is responsible for initiating and terminating
oscillations in \pft~concentration. The strength of the DNA damage signal
modulates the number of oscillatory waves of \pft~and \Mdm2~but not the
frequency or amplitude of oscillations--a result that recapitulates
experimental findings. A critical finding was that the phosphorylated form
of \nbs1, a member of the DNA break repair complex Mre11-Rad50-\nbs1 (MRN),
must augment the activity of \atm~kinase. While there is in vitro support
for this assumption, this activity appears essential for \pft~dynamics. The
model provides several predictions concerning, among others, degradation of
the phosphorylated form of \pft, the rate of DNA repair as a function of
DNA damage, the sensitivity of \pft~oscillation to transcription rates of
SIAH, $\beta$-CATENIN and ARF, and the hysteretic behavior of active
\atm~kinase levels with respect to the DNA damage signal.
\end{abstract}


\pagebreak

\section*{}
The \pft~protein is a transcription factor that is present in most higher
eukaryotes. It regulates the expression of a series of genes that mediates
cell cycle arrest and apoptosis in response to DNA damage and extracellular
signals~\cite{levine}. \pft~gene is often mutated in tumor cells
~\cite{ronai, malkin}. It is necessary for the elimination of potentially
cancerous cells by triggering cell death~\cite{levine2} and is thought to
be a major tumor suppressor gene~\cite{oren,lowe}. \pft~protein is subject
to a series of post translational modifications that modulate its kinase
activity~\cite{levine}. A simple view is that \pft~ is present in at least
two major forms, an unphosphorylated form that is thought to be inactive as
a transcription factor,  which is rapidly turned over by proteolysis and a
more stable phosphorylated form. Phosphorylation occurs at a number of
amino acid residues on the protein by the activity of protein kinases, and
the phosphate groups are in turn removed by the activity of several
phosphatases that are dynamically associated with \pft.

A phosphorylated form of \pft, \pftp, is the active transcription
factor that binds to a number of target DNA sites that regulate
the expression of several genes involved in cell cycle checkpoint
arrest and programmed cell death, respectively~\cite{kohn}.
Additionally, \pftp~ induces transcription of the gene {\it p53AIP1}, whose
product in the cytoplasm leads to the
release of cytochrome c protein from mitochondria, which in turn
activates programmed cell death directly by activating the
cytoplasmic protease cascade~\cite{matsuda}.

Ionizing radiation causes DNA double strand breaks that triggers the
binding of the sensor protein kinase \atm~\cite{mckinnon}to the nascent DNA
ends~\cite{tanya}. DNA binding triggers the kinase activity of
\atm,  which then phosphorylates preexisting \pft~proteins to
\pftp~\cite{abraham,kastan}. Unphosphorylated \pft~ is the
substrate of \Mdm, a ubiquitin E3 ligase~\cite{grossman}.
Once ubiquitinated, \pftp~protein is proteolytically degraded by
the proteosome complex. By contrast, \pftp~is more
resistant to ubiquitination, therefore is more stable. Hence
\atm~activation leads to the accumulation of \pftp. Nuclear
localized \pftp~binds to the regulatory sequences of several genes
among which MDM2 is one. This binding increases the transcription
of MDM2 gene. Thus an increase in \pftp~concentration in the
nucleus increases the synthesis of \Mdm~RNA and protein. The
increase in \Mdm~protein concentration is expected to destabilize any
remaining unphosphorylated \pft~by ubiquitination and
proteasome-mediated degradation.  Using this negative feedback,
\pft~regulates its own concentration in the cell~\cite{levine2}.

DNA double strand breaks induced by ionizing radiation are thought to cause
certain topological changes in the chromatin structure such that the
inactive \atm~kinase homodimer molecules associated with the chromatin
become activated~\cite{kastan}. One of the two \atm~kinase subunits
phosphorylates the other at the serine-1081 residue leading to reciprocal
phosphorylation of the two \atm~kinase subunits. The dimer is thought to
dissociate and the phosphorylated monomers have three central functions:
First, phosphorylation of nuclear-localized \pft~to \pftp; second,
phosphorylation of Brca1 and Brca2 proteins to activate cell-cycle
checkpoint arrest and blockage of DNA replication fork movement; and third,
recruitment of the MRN (Mre11-Rad50-\nbs) DNA repair complex to the nascent
DNA ends by protein-protein interaction while simultaneously activating the
function of the MRN complex by phosphorylating \nbs~to \nbsp~\cite{tanya1,
tanya2}.  The activated MRN complex begins to repair the broken DNA either
by nonhomologous end joining or by recombinational repair using the
unbroken homolog as a template. \pftp~directly recruits RAD51 protein at
the MRN complex by protein-protein interactions~\cite{buchhop} where RAD51
constrains the recombinational repair of DNA by the non-crossover mode.
This mode of DNA repair protects the genome from somatic aneuploidy that is
often an early event during cancer formation~\cite{duensing}. Yet another
effect of \pftp~is that it slowly accumulates in the cytoplasm where it
causes the mitochondrial membrane to leak cytochrome c protein
fragments~\cite{matsuda}. Accumulating cytochrome c in the cytoplasm is
ultimately responsible for inactivating the anti-apoptotic proteins
Bcl2/Bax, which leads to apoptosis. However, the relative rates of
cytoplasmic \pftp~accumulation and its nuclear effects in DNA repair
determine the fate of the cell: repair of DNA damage or death. Thus, the
levels of \pft~protein and its phosphorylated form are crucial determinants
of cell fate.

Recently Lahav et al.~\cite{lahav} measured intracellular
concentration of total \pft~and \Mdm~proteins by fusing their
coding sequences to fluorescent reporter domains. Examination of
single mouse cells following treatment with ionizing radiation
that produces DNA double stranded breaks (DSB) revealed that
\pft~and \Mdm~protein concentrations in singles cells oscillate in
response to DNA damage. The period and the amplitude of these
oscillations remained constant over a range of radiation doses,
but the number of pulses was proportional to the radiation dose.
Lahav et al.~\cite{lahav} proposed that the system behaves as a
``digital oscillator''.

In response, Tyson~\cite{tyson1} proposed that following DNA damage the
steady state \pft~concentration passes through a Hopf
bifurcation~\cite{strogatz,kuznetsov} and begins to oscillate. Once the
damage is corrected the system is pulled back from the oscillatory region
to its original steady state. This idea was further expanded in an explicit
model that assumed regulation of \pft~degradation by \Mdm~and \pft~as a
transcription factor for \Mdm~RNA synthesis~\cite{tyson2}. A crucial aspect
of this model is a positive feedback loop by which \pft~inhibits the
nuclear localization of \Mdm~into the nucleus~\cite{mayo}, thereby
increasing the \pft~levels.

Another approach to modeling the \pft~dynamics makes explicit use of delays
in the system corresponding to the time it takes for transcription and
translation of the \Mdm~protein~\cite{monk}~\footnote{ Tiana, G., Jensen,
M.H., \& Sneppen, K., (2002) {\it arXiv:cond-mat/0207236v1}}. Oscillations
are possible by including a sufficiently long delay, as for example in the
case of a hypothetical intermediate step introduced by Alon et
al.~\cite{alon}.

While previous modeling approaches explored the conditions that could in
principle generate oscillations in \pft~and \Mdm~levels, these models did not
specifically address explicit molecular mechanisms of DNA break repair or
control of \pft~activity. Here, in a significant departure from previous studies, we explicitly
model changes in unphosphorylated and phosphorylated versions of
\pft~in response to DNA damage. We ignore the nuclear transport of
\pft~because the time-scale of nuclear localization (minutes) is
fast compared to that of the observed oscillatory dynamics
(hours). We include in the model a specific pathway, that includes
the regulation of \Mdm~through \pftp~using \arf (p19),
$\beta\mbox{-catenin}$, and \siah~\cite{levine2}. We also explicitly
model the activation of the \atm~kinase, which phosphorylates \pft~as a switch.

The model reproduces the experimentally observed oscillatory dynamics in
\pft~and \Mdm~protein concentrations and generates a range of novel
molecular predictions that can be tested by future experiments. These
predictions include: the expected degradation profile of \pftp, the rate at
which DNA is repaired is proportional to the degree of DNA damage, the
sensitivity of \pftp~oscillations to the transcription rates of several
genes including SIAH, $\beta$-CATENIN and ARF, and, finally, specific
predictions on the hysteretic behavior of the \atm~kinase switch.

\section*{Methods}

All simulations were carried out using the Systems Biology Workbench
(SBW/BioSPICE) tools~\cite{omics}: the network designer, JDesigner, the
simulation engine Jarnac~\cite{jarnac}. Bifurcation diagrams were computed
using SBW with an interface to MATLAB~\cite{cameron}, and a bifurcation
discovery tool~\cite{chickarmane}. Bifurcation plots were also computed and
cross checked using
Oscill8~\footnote{http://sourceforge.net/projects/oscill8}, an interactive
bifurcation software package which is linked to AUTO~\cite{auto}, and
SBW~\cite{omics}. In all our simulations the species concentrations are
regarded as dimensionless, whereas the kinetic constants have dimensions of
inverse time, with dimensionless Michaelis-Menten constants. All models are
available as standard SBML or Jarnac scripts. In the sensitivity analysis
of a network, the control coefficients $C^{i}_{j}$ are defined as,
$C^{i}_{j}=\frac{p_j}{S_i} \frac{\partial S_i}{\partial p_j}$, where $S_i$,
$p_j$ are the species, and the parameters respectively.

\section*{Results}
\subsection*{The Model}
\begin{figure}[t]
\begin{center}
\includegraphics[scale=0.5]{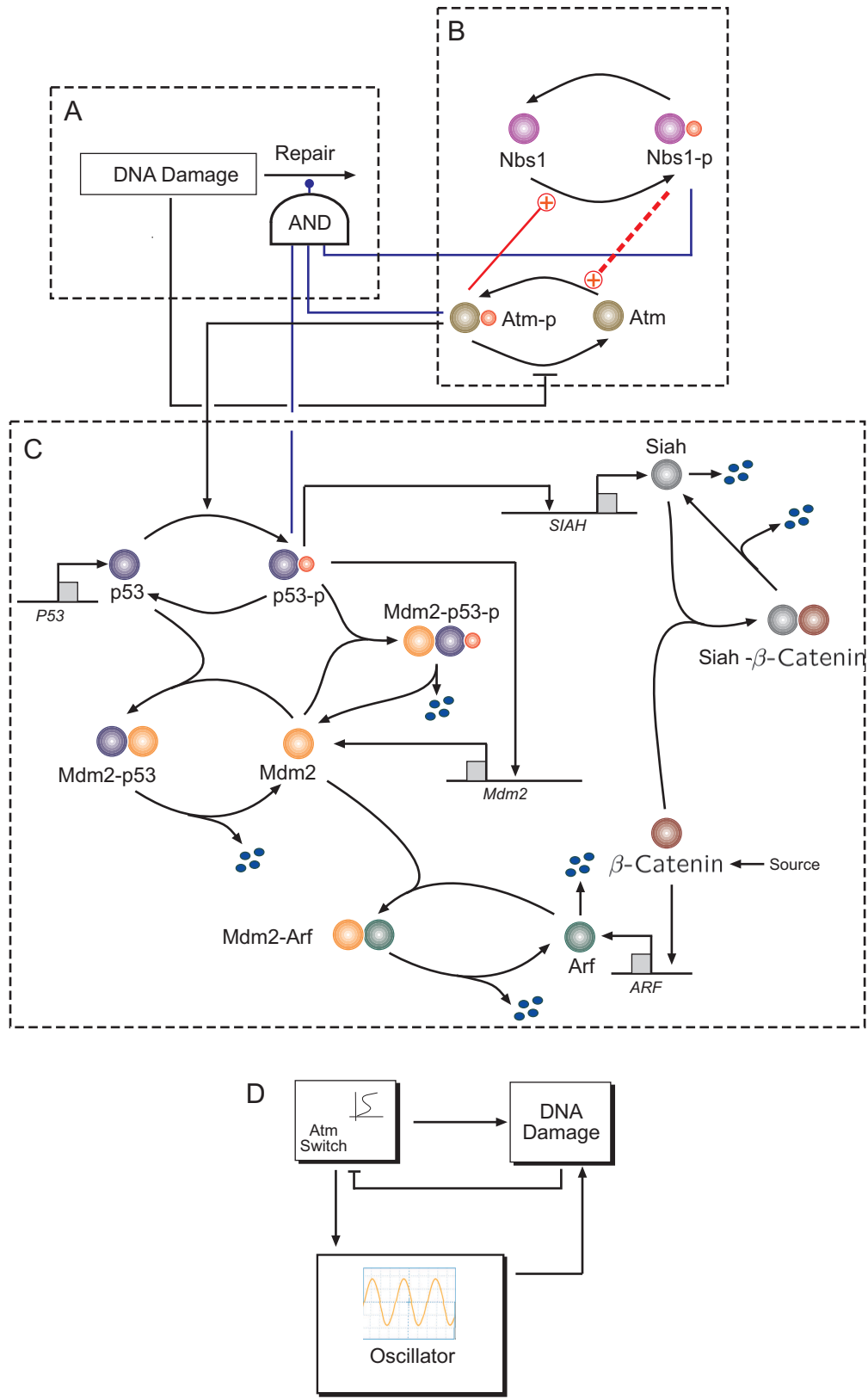}
\end{center}
\caption{The core regulatory network. The three modules labeled as
A, B and C, refer to the DNA damage
repair, the \atm~kinase switch, and the \pft~oscillator respectively.
These are described in detail in the text. Panel D presents a cartoon of
the regulation for the integrated \pft~network.  }
\end{figure}

We modularize the network~\cite{sauro} into three distinct dynamically
different pieces: a switch describing the \atm~kinase,  a subnetwork that
describes the DNA damage repair, and an oscillator that includes \pft~and
\Mdm~~(Fig 1). DNA damage turns on the \atm~kinase switch which then turns
on the oscillator. The oscillations activate and maintain the repair
process. As the damage is corrected, the switch is shut off; this in turn
switches off the oscillator~~(see Fig 6).

\subsection*{Module A: DNA Repair}

We assume that when DNA is damaged a signal \dam, which is proportional to
the damage, is generated. Repair of DNA breaks decreases \dam. In reality,
\dam~is equivalent to the number of nascent double stranded DNA ends
created due to the interaction of ionizing radiation with DNA. In response
to the DNA damage, \atm~is phosphorylated to \atmp, and the activated
\atmp~kinase in turn phosphorylates a host of target proteins, including
\nbs, Brca1, Brca2 and \pft. Brca1 and Brca2 block DNA replication and cell
cycle. The phosphorylated form \nbsp~ recruits Mre11 and Rad50, and the
complex of three proteins (the MRN complex) binds the broken DNA to
initiate repair-recombination. Hence the repair process begins once
concentrations of \atmp~kinase, \nbs~kinase and \pftp~begin to grow. The
rate of repair is thus described by

\begin{equation}
\frac{d [\dam]}{dt} = -\alpha_D [\pftp][\nbsp][\atmp][\dam],
\end{equation}

where $\alpha_D=2\times10^{-4}$, is a rate constant of the decay
in damage signal, which can be calculated by measuring the change
in the number of DNA ends per genome. The bracketed variables
refer to the concentration of the various species, in
dimensionless form. For simplicity we assume that all double
strand break (DSB) repair is dependent on \pft; this is clearly
untrue because more error prone \pft~independent repair pathways
exist.

\subsection*{Module B: \atm~Switch }

\atm~proteins are normally sequestered in the form of dimers/multimers, and
are thought to undergo autophosphorylation ~\cite{kastan,kurz}. A class of
phosphatases, PP-2A, are known to dephosphorylate  \atmp~\cite{khanna}.
When a DSB occurs, two events happen very quickly. The first is that the
phosphatases dissociate from \atm~multimers and this results in an
increased phosphorylation rate, and the second is that the \atm~dimers
dissociate to activated monomers. It has been hypothesized that ``turning
off'' the suppressive effect of the phosphatases on the autophosphorylation
of \atm~allows rapid signal transduction of the DNA damage
signal~\cite{khanna}. Upon DNA damage, \atmp~accumulates rapidly in the
nucleus~\cite{kastan}. We assume that the effect of ionizing radiation is
to reduce the ability of the phosphatases to bind to \atm. This reduction
in the phosphatase activity following DNA damage could be thought to occur
due to competition among \atmp~ and the signal \dam~ to bind to the
phosphatases. However, we do not explicitly model the competitive
 inhibition but use a mathematical rate law that is derived from such an
assumption. It is known that \atmp~phosphorylates \nbs, which is part of
the MRN complex, and that the MRN complex further increases the activity of
\atm~\cite{tanya1, tanya2}. We take this as a cue to modelling the
interaction between \atm~and \nbs~as a positive feedback system. As such,
active \atm~(i.e. phosphorylated \atm) will activate \nbs~in the MRN
complex (through phosphorylation), leading to further activity in \atm.
This explains why \atmp~levels increase rapidly in the cell upon DNA
damage. Thus, \nbsp~positively regulates (thick dotted line in  Fig. 1,
Module B) activity of \atm~by causing its phosphorylation. Alternatively,
one can regard this effect to be due to a further activation of \atmp~ by
\nbsp; for simplicity we do not introduce a new variable for the ``active''
form of \atmp. We do not know the exact process by which this may occur,
but find that such a process is required by the model to be able to predict
the experimental observations of the \pft~dynamics (see later).

Referring to Fig.~1, Module B, the equations (see supplementary materials for
reaction scheme) for \atm, and \nbs~are,

\begin{eqnarray}
\frac{d [\atmp]}{dt} &=& \frac{[\atm][\nbsp]}{k_{1} + [\atm]}
+\alpha -\frac{[\atmp]}{k_{2} +[ \atmp]}
\left(\frac{\alpha_s}{k_{0}+[\dam]}\right),\\\nonumber \frac{d
[\nbsp]}{dt} &=& \frac{[\nbs][\atmp]}{k_{4} + [\nbs]} +\beta
-\frac{[\nbsp]}{k_{3} + [\nbsp]}\\\nonumber
\end{eqnarray}

where the assumption of the positive feedback (marked as the dotted line in
Fig. 1 Module B) is given by the first term in the rate law for \atmp.
$\alpha$, $\beta$, are basal rates of phosphorylation due to stochastic
fluctuations in \atm~and \nbs~kinase activities, respectively. In the above
equations, the rate constants are in units of inverse time, $\alpha_s$ is a
scaling parameter, and the Michaelis-Menten constants ($k_{1}, k_{2},
k_{3}, k_{4}, k_{0}$) are in dimensionless units.The parameter values for
these equations are reported in supplementary materials.
\begin{figure}[h]
\begin{center}
\includegraphics[scale=0.5]{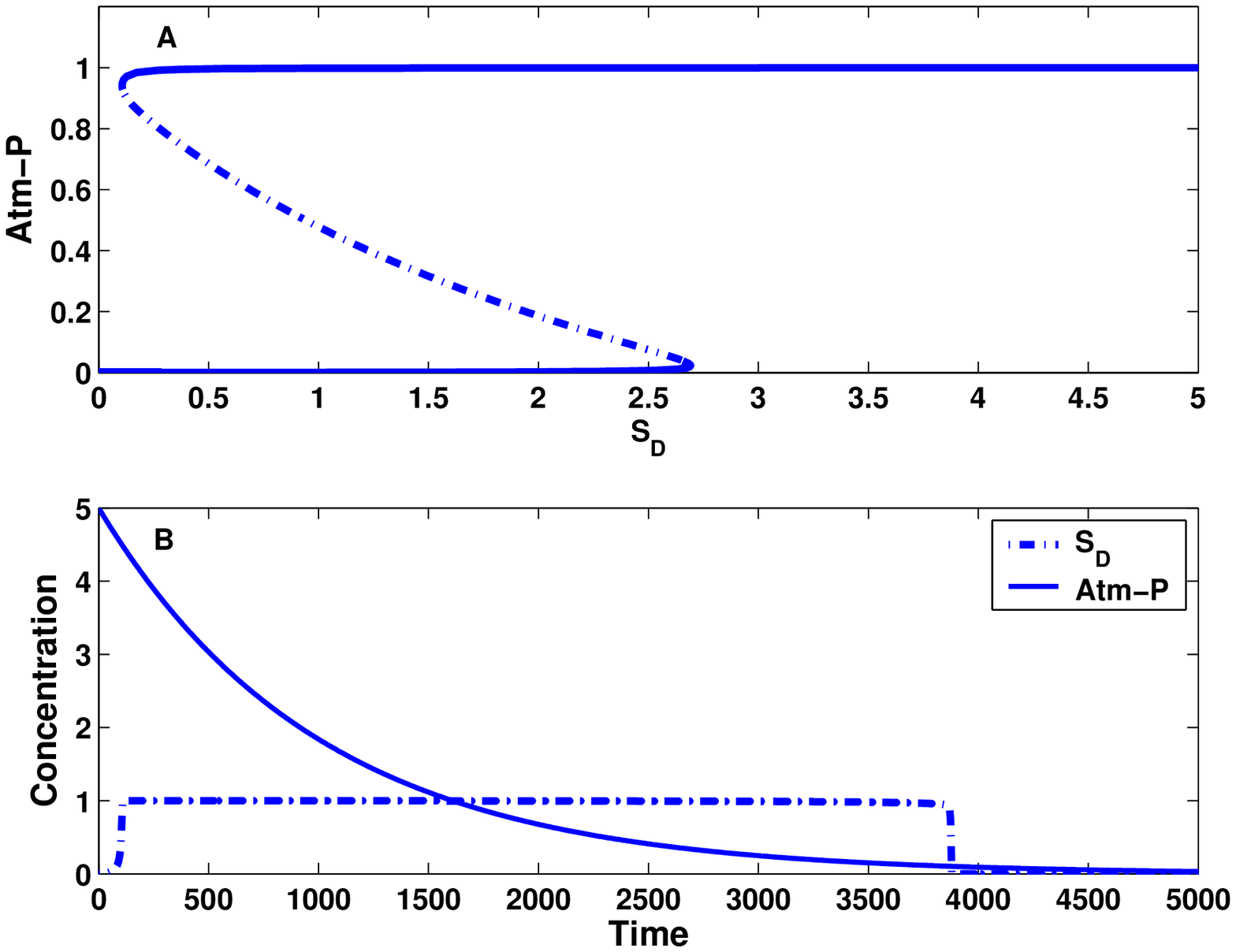}
\end{center}
\caption{Bifurcation diagram and time series plot for the
\atm~kinase activity exhibiting bistability. The upper plot shows the steady
state values of \atmp~as a function of the DNA damage signal \dam.
The thick lines indicate the stable points, whereas the dotted
lines indicate the unstable points. For \dam~$
> 2.7$, the system switches from  a low value of \atmp, to a high
value of \atmp~(a similar bistable effect is observed for \nbsp). The lower
plot shows the behavior of \atmp~as a function of time. In this simulation,
\dam~was modelled to be an exponentially decreasing function of time, and
hence used as a parameter in Eq. 2. Initially the damage causes \atmp~to
switch to a high level ($\simeq 1$), and then subsequently, as
\dam~decreases and crosses a certain threshold, the \atmp~ switches back to
a low value. The plot shows that as damage crosses $\simeq 0.08$, the
switch is turned {\bf OFF}, and one obtains a sharp crossover.}
\end{figure}

Using the above equations the steady state values of \atmp~reached by the
system as a function of the DNA damage, \dam, was computed in Fig.~2A. The
system shows bistability, i.e. two possible steady states for a range of
\dam.  For large values of damage (\dam~$ > 2.7$), \atmp~$ = 1$, and the
switch is in the {\bf ON} state. As the damage is corrected, and drops
below $\simeq 0.08$, the switch is turned {\bf OFF}. The plot for \nbsp~is
very similar, and this is due to the nature of the positive feedbacks that
\atm~and \nbs~have on each other. There are several examples where positive
feedback can give rise to bistable, switch-like, behavior~\cite{ferrell1,
ferrell2, chen}. The system can therefore move between two states, a low
state defined by low values of \atm~and \nbs, and a high state defined by
high values of \atmp~and \nbsp. As shown in Fig.~2B, as the damage
decreases below $\simeq 0.08$, the switch is turned to the {\bf OFF} state,
and \atmp~levels drop off rapidly.

In the present model the positive feedback between \atmp~and \nbsp~has
significant effects on the dynamics of the integrated system. Neglecting
the positive feedback still gives a quick response in \atmp~levels to DNA
damage, but as the DNA is repaired, the \atmp~ levels do not fall off
abruptly, but gradually decrease (supplementary materials). For the
integrated model to produce the observed~\cite{lahav} results that describe
the experimental observations it is necessary to have a switch-like
behavior in \atmp~levels that in turn requires the stated positive
feedback.

\subsection*{Module C: \pft~Oscillator:}

To determine whether the known regulatory network involving \pft~and
\Mdm~is sufficient to reproduce their observed dynamics, we proceed as
follows. The central oscillator is composed of the negative feedback of
\Mdm~on \pft~stability through ubiquitination, and the dependence of
\Mdm~expression on \pftp. The latter is produced by \atmp~  in response to
the DNA damage. The dephosphorylation of \pftp~is described by a
Michaelis-Menten scheme, and is assumed to occur at a very small rate
compared to the phosphorylation rate. We assume small first order
degradation rates for both \pft~and \pftp. \pftp~is a more stable form of
\pft, and hence its rate of degradation is smaller than the degradation
rate for \pft. We model the degradation of \pftp~as a function of \Mdm;
this is done to match qualitatively the simulation results of our model to
the experimental results (see below). Furthermore, \pft~indirectly
regulates the amount of \Mdm~ that is available for binding to \pft~
through the former's transcription factor activity on the SIAH
gene~\cite{levine2}. The \siah~protein is a ubiquitin ligase, and enhances
the degradation of $\beta\mbox{-catenin}$. $\beta\mbox{-catenin}$, which is
produced constitutively, regulates the transcription of the \arf~gene. The
\arf~protein binds to \Mdm, sequestering the latter from its substrates and
also promoting the latter's degradation such that increasing concentration
of \arf~should increase the \pft~level~\footnote{Sionov, R.V., Hayon, I.L.,
\& Haupt,Y.,(2000-2005) The regulation of \pft~ growth
supression,http://www.ncbi.nlm.nih.gov/books/bv.fcgi?rid=eurekah.chapter.11736.}.
The following differential equations describe the dynamics of module C in
Fig. 1, and were generated automatically by SBW from SBML~\cite{omics}.

\begin{eqnarray}
\frac{d[\pft]}{dt} &=& c_0 -  \frac {[\atmp][\pft] }{k_{p}+ [\pft]} - con
+ P_h \frac{[\pftp]}{(k_{dep}+[\pftp])}\\ \nonumber &-& k_{1p}
[\pft] [\Mdm] - k_{2p} [\pft],\\ \nonumber \frac{d[\pftp]}{dt} &=&
 \frac {[\atmp][\pft] }{k_p+ [\pft]} + con
- k_{p3} [\Mdm] [\pftp] \\
\nonumber &-&P_h~\frac{[\pftp]}{(k_{dep}+[\pftp])} -k_d
[\pftp],\\\nonumber \frac{d[\Mdmpft]}{dt} &=& k_{1p} [\pft] [\Mdm] -
k_{c1} [\Mdmpft],\\\nonumber
 \frac{d[\Mdmpftp]}{dt} &=&  k_{p3} [\pftp]
[\Mdm] - k_{c2} [\Mdmpftp],\\\nonumber
\\\nonumber
\frac{d[\Mdm]}{dt} &=& P_1 \frac {[\pftp]^4}{k_m+ [\pftp]^4} - k_{1p}
[\Mdm] [\pft] -k_{4a} [\Mdm] [\arf]-k_5 [\Mdm]\\\nonumber &-& k_{p3}
[\Mdm] [\pftp] + k_{c2} [\Mdmpftp]+ k_{c1}[\Mdmpft],~\\\nonumber
\frac{d[\Mdmarf]}{dt} &=& k_{4a} [\Mdm][\arf]-k_6
[\Mdmarf],~\\\nonumber \frac{d[\arf]}{dt} &=& p_0
[\beta\mbox{-catenin}]-k_{4a} [\Mdm][\arf] -k_7 [\arf] +k_6
[\Mdmarf],\\\nonumber \frac{d[\siah]}{dt} & =&  p_3 \frac
{[\pftp]^4 }{k_s+ [\pftp]^4}+k_{c3} [\beta\mbox{-catenin-}\siah]
-k_{c4} [\beta\mbox{-catenin}] [\siah]-k_{si} [\siah],\\\nonumber
\frac{d[\beta\mbox{-catenin}}{dt} &=& p_{1p} -k_8
[\beta\mbox{-catenin}] - k_{c4} [\beta\mbox{-catenin}]
[\siah],~\\\nonumber \frac{d[\beta\mbox{-catenin-}\siah]}{dt} &=&
k_{c4} [\beta\mbox{-catenin}] [\siah]-k_{c3}
[\beta\mbox{-catenin-}\siah]
\\\nonumber
\end{eqnarray}

The parameter values for these equations are given in supplementary
materials. The transcriptionally active \pftp~proteins form a
tetramer~\cite{lee}, and hence the transcription rates for MDM2 and ARF
genes are described by a Hill coefficient of 4. The concentration of \atmp,
which determines the amount of phosphorylated \pft, is treated as a
parameter in these equations.

\begin{figure}[hbtp]
  \vspace{9pt}
\centerline{\hbox{ \hspace{0.0in}
\includegraphics[scale=0.45]{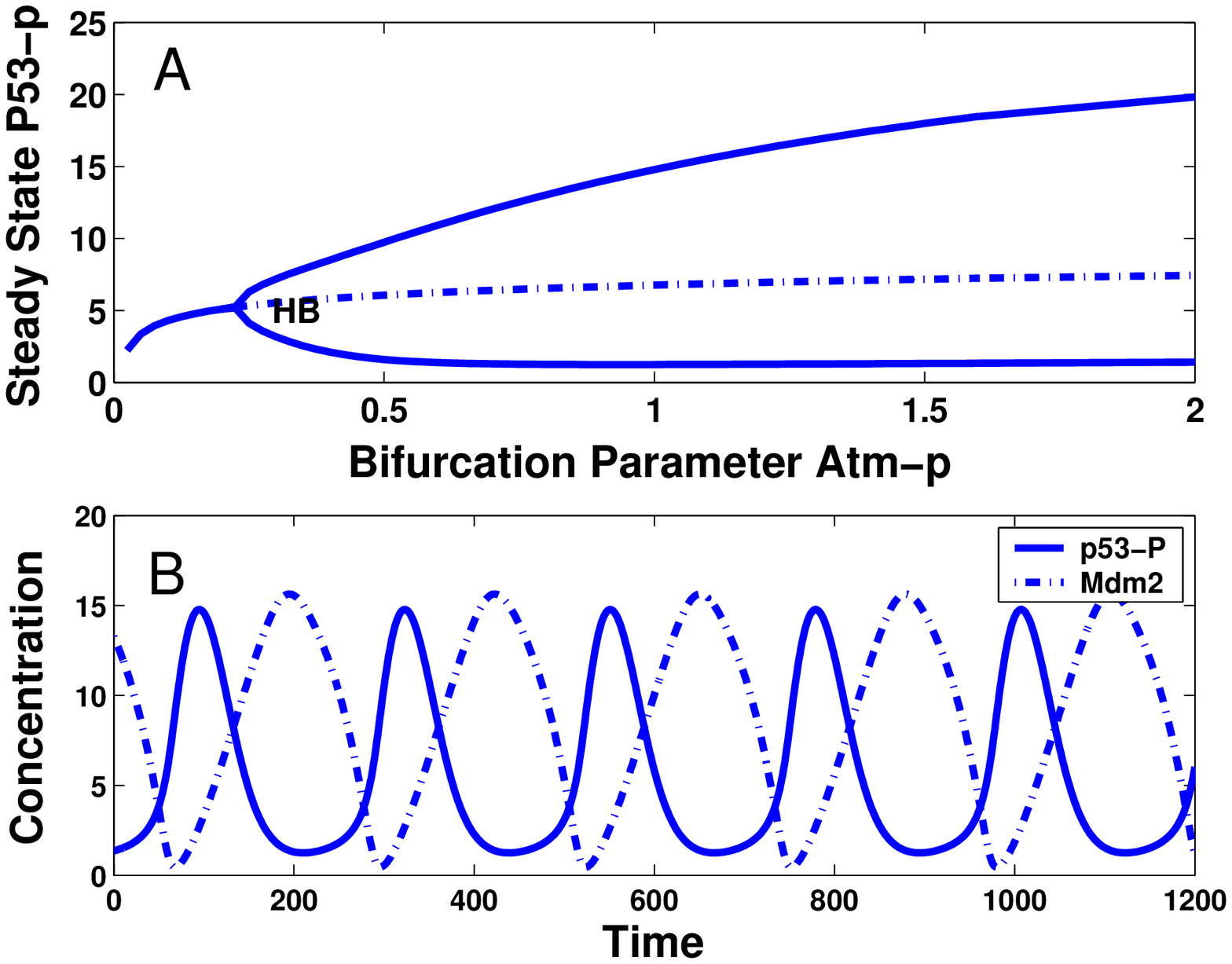}
    \hspace{0.2in}
    }
  }
  \vspace{9pt}
\caption{\pft~and \Mdm~dynamics.  Panel A shows the steady state value of
\pftp~as a function of \atmp~which shows a supercritical Hopf bifurcation
(indicated by point HB) that occurs at $\atmp \simeq 0.22$. At this point
the system develops stable oscillations, where the amplitudes of the
oscillations increase with increasing \atmp. The limits of the oscillations
are marked by the thick lines. The dashed line after the point marked HB
indicates the unstable branch. In panel B, we display
the time series of \pftp~and \Mdm, which can be seen to be out of phase. The
overlap between the time series, occurs due to the activation-inhibition loop described
in the model. For the plotted time series, the value of $\atmp=1.0$.
}
\end{figure}

In Fig.~3A, we plot the steady state concentrations of \pftp~as a function
of the \atmp~concentration. \pftp~exhibits oscillations for $\atmp~> 0.22$,
i.e. in a region beyond the Hopf bifurcation. The kinetics for \pftp~and
\Mdm~concentrations, plotted in Fig.~3B for $\atmp=1$, show that they are
out of phase with each other. \pftp~is assumed to decay by two mechanisms:
a first order decay that is independent of \Mdm~concentration, and another
that is dependent on \Mdm~concentration. The second decay rate was assumed
because without this critical assumption, the amplitudes of
\pftp~oscillations do not fall to low levels (supplementary material) as
was observed in the experiment~\cite{lahav}.

A sensitivity analysis shows that if the transcription
rate of MDM2 is decreased, then this can be balanced by a slight
increase in the transcription rate of SIAH, and vice versa. The decrease
in SIAH transcription leads to a release of \Mdm~that was previously sequestered
by \arf, and this counteracts the decrease in \Mdm. The control coefficients, $C^{\pftp}_{P_1}$,
$C^{\pftp}_{P_s}$ are both negative, and are approximately of the
same order of magnitude(see Supplementary materials). Increasing either of these two
transcription rates leads to a decrease in \pftp, and hence they are mutually compensatory.

\subsection*{Integrated Model}

The three modules are now combined into one integrated model. Eqs. 1, 2 and
3 are simulated with appropriate initial conditions to obtain the
concentrations of the various species as a function of time. The initial
conditions are the steady state concentrations of \pft, \pftp, \Mdm, \arf,
$\beta\mbox{-catenin}$, \siah, and the complexes. These steady state values
are obtained by solving the rate equations for \dam~$=0$. The DNA breaks
are assumed to occur at $t=0$. Although the DNA breaks occur over a finite
time interval, we assume that this is short compared to the period over
which repair occurs. The DNA damage turns the \atm~switch {\bf ON}, which
rapidly makes \atmp~$ = 1$, and \nbsp~$= 1$ (see Fig.~2B). Since \atmp~$>
0.22$ (from Fig.~3A), the system crosses the Hopf Bifurcation point, and
enters the oscillatory regime. \pftp~begins to oscillate, and \pftp,
\atmp~and \nbsp~(through their respective actions not explicitly modelled
here) begin to correct the damage (Eq. 1). As DNA repair proceeds,
\dam~decreases. At a certain threshold, \atmp~falls sharply back to its
original value of zero (similarly for \nbsp). The \atm~switch is shut {\bf
OFF}, and the system exits out of the oscillatory zone back to its original
stable state. In Fig.~4A, we display the concentration of \atmp, which
rapidly rises to $\simeq 1$, remains at this value for a while, and finally
rapidly declines as \dam~decreases below a certain threshold. Also shown is
the dynamics of \pft~which falls from its steady state value close to
$\simeq 2$, and oscillates with a small amplitude, but eventually rises to
its original steady state value, as the repair is completed. All the other
species described in Fig.~1 Module C also oscillate (i.e.~\siah,
$\beta\mbox{-catenin}$, \arf~etc). In Figs.~4 (B-D), we plot the time
series values of \dam, \pftp~and \Mdm~ for different initial values of DNA
damage.

\begin{figure}[hbtp]
  \vspace{9pt}
\centerline{\hbox{ \hspace{0.0in}
\includegraphics[scale=0.45]{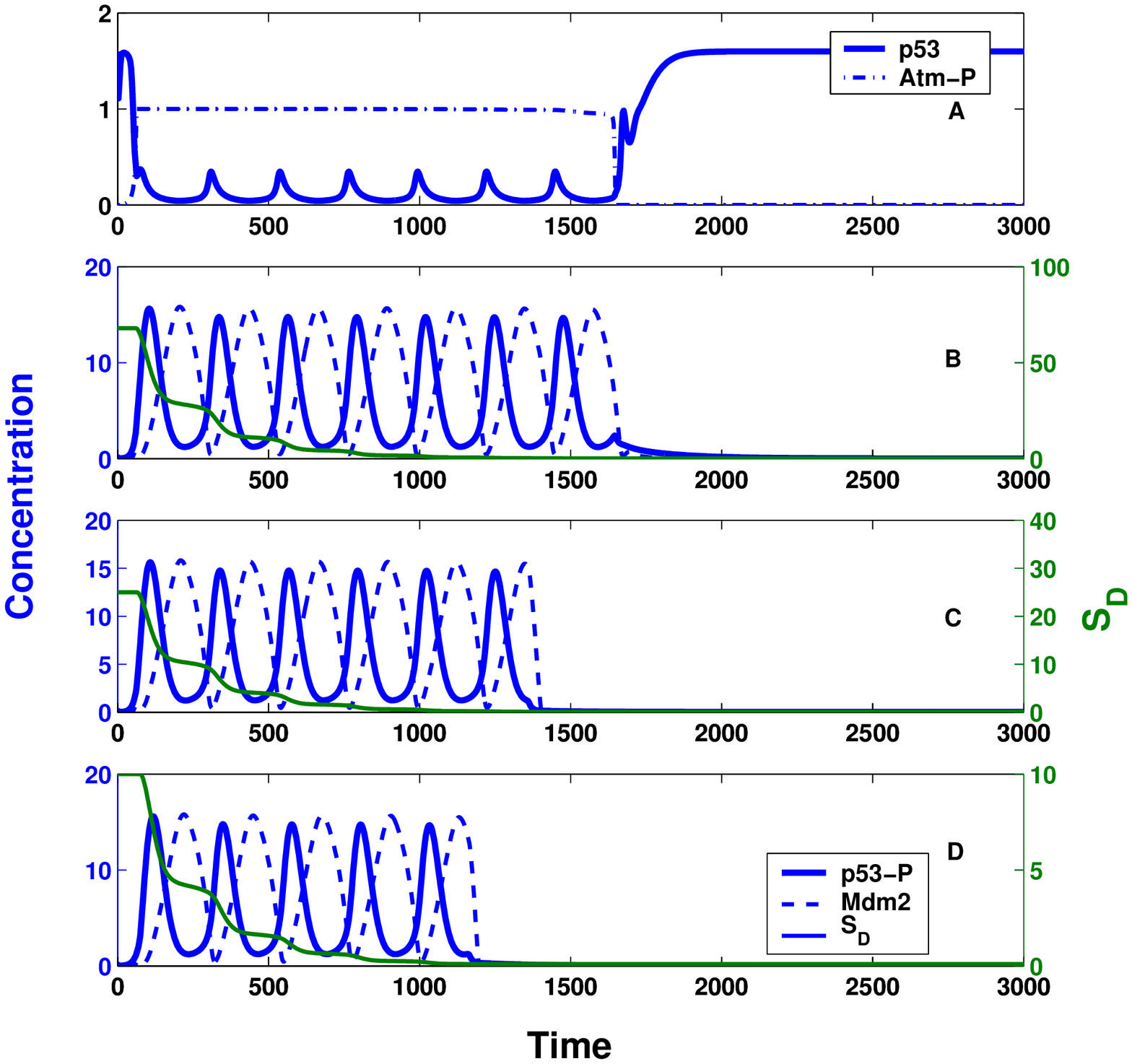}
    }
  }
  \vspace{9pt}
\caption{Effects of DNA breaks on \pft~and \Mdm~dynamics.
Panel A:Time-series plots for the concentrations of time courses for \atmp~and \pft, for
$\dam=68$, at $t=0$.
Panels B-D:Time-series plots for
\pftp, \Mdm~and \dam~ as a function of various initial damage levels,
$S_D(t=0)=68, 25, 10$. The number of pulses increases with damage,
while retaining their amplitude and frequency (in our model, the
number of pulses is not a linear function of the damage. If
however, we assume a Michaelis-Menten type of rate law for the
damage such as assumed in~\cite{tyson2}, i.e the rate of change of
damage is, $\propto -[\pftp] \frac{\dam}{k_{dam} +\dam}$, then
for small $k_{dam}$, the rate of decay of damage( for \dam$> 1$)
is proportional to the concentration of \pftp~(since \nbsp,
\atmp$\simeq 1$). From this it follows that the number of \pftp~pulses required for repair
will vary approximately linearly with respect to the amount of initial DNA damage.)}
\end{figure}

Note that the number of pulses increases with damage; however, the shape
and frequency of the pulses do not change. This is consistent with the
experimental observations that describe the ``digital'' behavior of
\pft~oscillations~\cite{lahav}. On DNA damage the system is moved into the
oscillatory zone rapidly, by the \atmp~switch. Once inside the oscillatory
zone, the switch is {\bf ON} at a fixed value of $\atm~=1$, which fixes the
amplitude of the oscillations. Similarly when the switch turns {\bf OFF}
the system moves out of the zone rapidly. Hence even though in this case
the system exhibits a supercritical Hopf bifurcation, for which the
amplitude of oscillations vary slowly with the value of \atmp, the rapid
movements into and out of the oscillatory regions give rise to fixed
amplitude and frequency. An alternative model, as proposed in~\cite{tyson2}
is to employ a subcritical Hopf bifurcation which has the natural property
to generate large amplitude oscillations with only small parameter changes.

\section*{Discussion}

The model presented here describes the dynamics of a gene-regulatory
network involved in DNA damage repair. We describe the model as three
interacting modules. The \pft, \Mdm, \siah, $\beta\mbox{-catenin}$, and
\arf~together represent the oscillator. The \atm, \nbs~dynamics presents a
switch. The repair process is active as long as \pftp, \atmp~and~\nbsp~are
present. By modularizing the network we identified the key dynamical
assertions (Fig.~1D). The \atm, \nbs~switch is activated by DNA damage; the
switch turns on \atmp~and \pft~ is phosphorylated; \pftp~then begins to
oscillate due to its interaction with \Mdm~ (also through \arf, \siah, and
$\beta\mbox{-catenin}$). The damage begins to be repaired because \pftp,
\atmp~and \nbsp~are abundant. The number of pulses of \pftp~is proportional
to the damage but the frequency and amplitude of the pulses remain the
same. As \dam~ decreases the \atmp~activity is switched {\bf OFF} and the
system comes back to its steady state. Our model therefore supports a
``digital'' oscillatory behavior.

An important point is the sensitivity of the dynamics of the integrated
model to parameter values. In Fig.~5 we plot on a logarithmic scale the
range of parameter values for which the model exhibits qualitatively
similar behavior.

\begin{figure}[h]
\begin{center}
\includegraphics[scale=0.5]{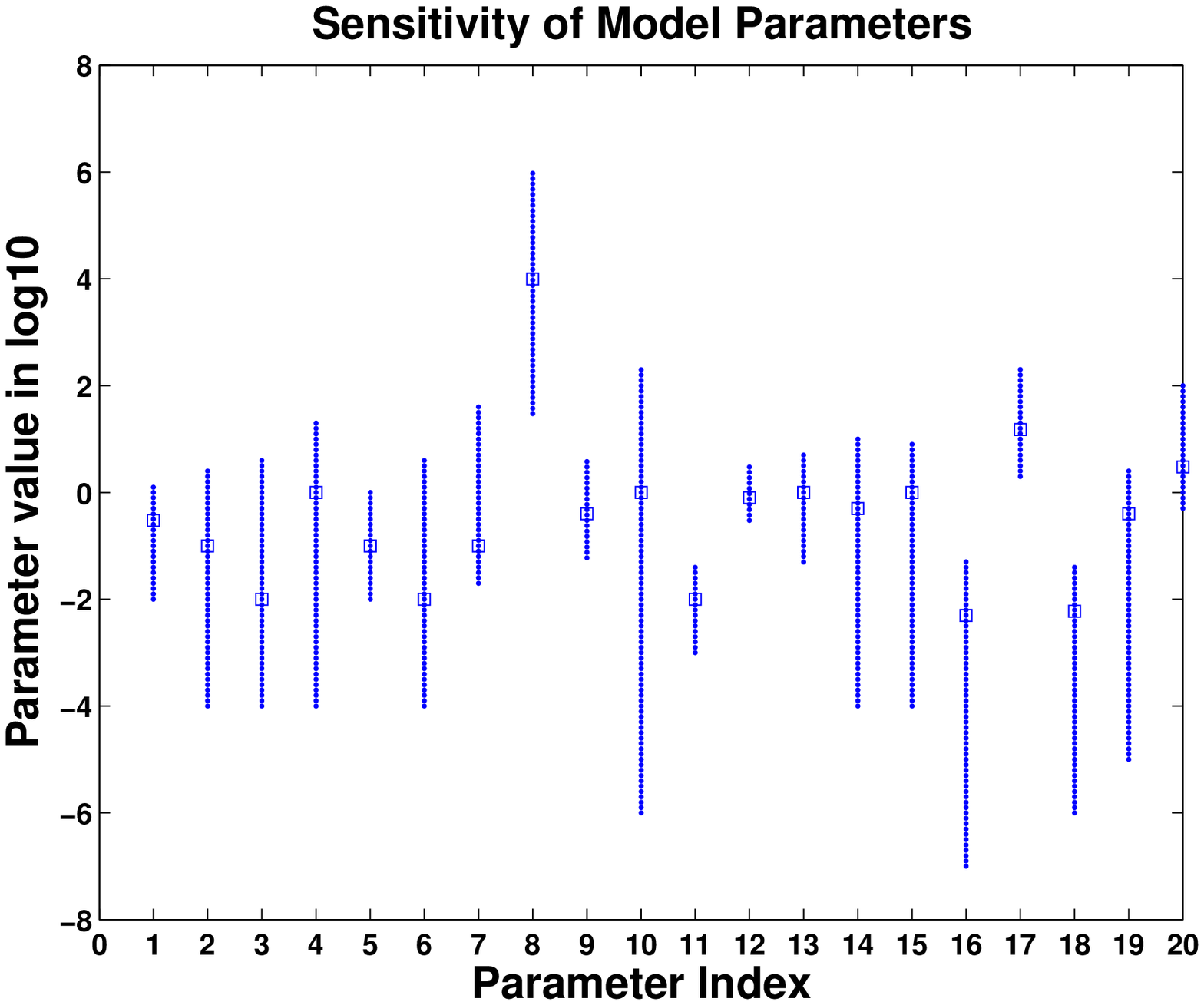}
\end{center}
\caption{Sensitivity of model parameters to the system dynamics. A plot of
the range of values of several parameters in logarithmic scale (base 10),
vs the parameter index. The parameters are numbered from 1-20, and are
listed below. The range (length of vertical line) indicates the set of
values for which the system exhibits similar dynamics. The square symbol in
each vertical line indicates the nominal value of the parameter in the
model about which the range is computed. A short line segment indicates
higher sensitivity to changes in the dynamics of the model, whereas a long
line segment indicates robustness. For example, the model is very robust to
changes in $k_m$ (index 10), which is the Michaelis-Menten constant for the
rate of production of \Mdm. Since the hill coefficient for this rate is 4,
the transcription rate is fairly independent of the value of $k_m$. The
model is sensitive to changes in $c_0$ (index 12), the rate of production
of \pft. The steady state concentration of \pft~is relatively low in the
cell, and this restricts the value of $c_0$ from assuming large values,
whereas extremely low values of $c_0$ are insufficient to provide enough
\pft, \pftp~to start the oscillations. The sensitivity plots were obtained
by a combination of bifurcation studies and simulations for each of the
following parameters in the same order as the ``Parameter Index'' in the
figure. $\alpha$, $k_{0}$, $k_{1}$, $k_{2}$, $k_{3}$, $k_4 $, $P_2$, $k_s$,
$P_1$, $k_p$, $P_b$, $c_0$, $p_1$, $k_d$, $k_{1p}$, $k_{p3}$, $k_{4a}$,
$k_d$, $k_p$, $k_{c4}$   }
\end{figure}

There are a few critical parameters that determine the period and the
amplitude of \pftp~oscillations. Specifically, the rates of transcription
of SIAH and ARF  have significant effects on the period. This is
intuitively obvious because these rates determine the rate at which \Mdm~is
released from sequestration due to binding with the \arf~protein.
\pftp~amplitude can be increased by changing the Michaelis-Menten constant
that determines the rate of phosphorylation (i.e $k_p$). Similarly it can
be decreased by increasing the transcription rate of MDM2 gene. These
observations suggest simple tests. For example, partial functional knockout
by siRNA could reduce the mRNA from any of the genes that form part of the
\pft~oscillator circuit. The model predicts that \pft~oscillations would be
affected by these partial knockouts.

The degradation of \pftp~is a largely unknown process. If we only assume a
simple first order rate of degradation proportional to \pftp~the results of
the simulations are inconsistent with experimental
observation~\cite{lahav}. In particular the troughs in the oscillations of
\pftp~do not reflect the low levels observed in the experiment. To
accommodate this discrepancy, we proposed an additional degradation route
which is a function of both \pftp~and \Mdm. Although the mechanism assumed
here is similar to ubiquitination, there may be another mechanism though an
intermediate gene product. With this addition to the model, the simulations
faithfully reproduce the experimental observations. Experiments to test
this alternative degradation routes should constitute removing \Mdm~and
measuring \pftp~degradation using antibodies specific to this isoform.

A crucial aspect of the model is the positive feedback regulation between
\atmp~and \nbsp. This leads to a bistable switch-like model for the
\atm~activity. Here we have not explicitly modelled the mechanism of
\atm~activation by DSB, which may or may not require MRN. However, the
positive feedback between \atm~and \nbs~is essential. Most recent in vivo
experiments provide additional support for the positive feedback activation
of \atm~by \nbs~\cite{diphil}. To verify the \atm~switch behavior, it is
necessary to determine the two damage thresholds that occur in the
bifurcation plot (Fig.~2A): The first is the minimum damage required to
throw the \atm~switch {\bf ON} (\dam~$\simeq 2.7$), and the second is the
damage level at which the \atm~switch turns {\bf OFF} (\dam~$\simeq 0.08$).
The first threshold can be measured if cells are subjected to increasing
doses of irradiation, and the damage level that turns \atmp~level {\bf ON}
is determined. Once the \atm~switch is {\bf ON}, we wait for the switch to
turn {\bf OFF} and at that point measure the existing DNA damage. The model
predicts that the time taken to correct a larger amount of damage will be
almost as long as that for a small amount of damage. This is because for
high \dam, \atmp~levels are high and the repair progresses relatively
rapidly, whereas for low \dam, the \atm~switch is {\bf OFF} and so
correction must occur through basal amounts of \atmp.

Transgenic mice with extra copies of \pft~genes are resistant to tumor
formation and age normally, suggesting that increased doses of \pft~gene
may allow normal regulation~\cite{garcia}. By contrast, mice with a
dominant negative form of \pft~gene, whose product is always {\bf ON} but
is not subject to \Mdm~action, are radiation resistant but age faster. We
predict that in these latter mice, \pft~protein levels do not
oscillate~\cite{tyner}.

We note (Fig.~6, curve B) that by adjusting $\alpha$ and $\beta$ (i.e.,
increasing the basal phosphorylation rates) it is possible to generate high
levels of \atmp~even for a low \dam. By this analysis, if the basal rates
of phosphorylation of \atm~could be increased then the time taken to
correct a small amount of damage could be decreased. This is testable by
decreasing the concentrations of phosphatases through the use of siRNA.
Furthermore, if $k_0$ is decreased, which corresponds to inefficient
phosphatase binding to \atm, the steady state values of \atmp~for small
\dam~ remain at a high level (Fig. 6, curve A). Therefore it would be
instructive to examine the possible existence of cell lines where the
levels of \pft~naturally oscillate even in the absence of much DNA damage.
We predict that, if such cell lines exist, mutations in these lines should
affect the binding of phosphatases to \atm.

\begin{figure}[h]
\begin{center}
\includegraphics[scale=0.5]{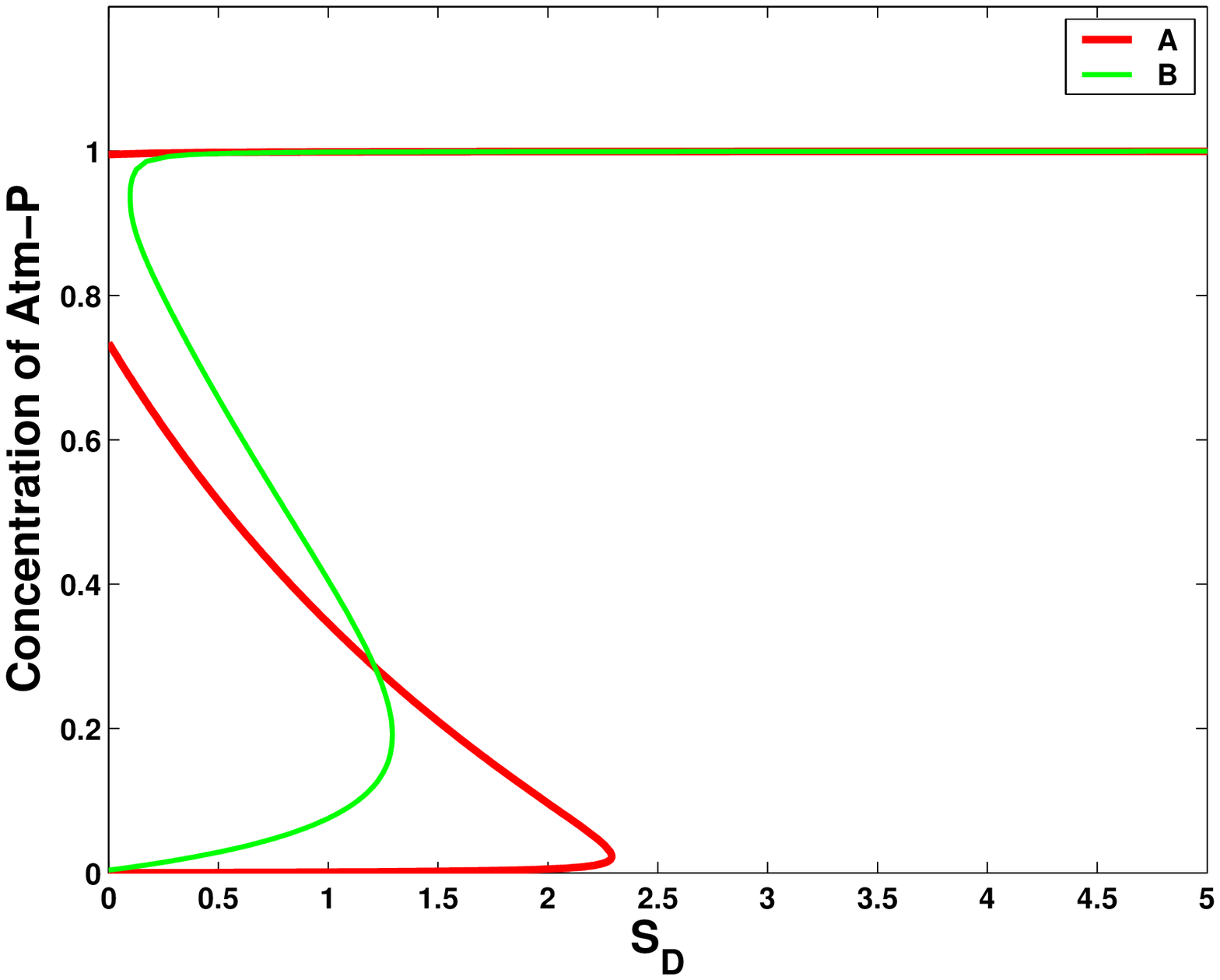}
\end{center}
\caption{Bifurcation diagram for the steady state values of
\atmp~as a function of \dam.
Curve A: A bifurcation plot, for modified Phosphatase activity, which is obtained by changing $k_0$.
The upper branch of the steady state values of \atmp~do not fall to zero, at low values of the DNA damage signal \dam,
which means that as the damage is corrected the \atm~switch is still in the {\bf ON} state, and hence \pft~oscillations
continue even when damage is corrected.
Curve B: The steady state values of \atmp~are reasonably higher at low values of the DNA damage signal \dam, as compared
to a similar plot in Fig.~2, panel A. These higher basal rates of \atmp~and \nbsp~would lead to faster DNA repair, for the same amount of
small DNA damage, when compared to the case considered in Fig.~2, panel A. This plot has been obtained by increasing the values of $\alpha, \beta$
which corresponds to higher basal levels of phosphorylation (see supplementary materials for parameter values)}
\end{figure}

In this work we have neglected the effects of stochastic fluctuations in
chemical species by assuming that the concentrations of all species are
large. In a future publication we will examine the effects of stochastic
noise on small concentration ranges.

\section*{Acknowledgement}
This work was supported by grants from the National Science Foundation (EIA
0205061 and FIBR 0527023 to A.R; 0432190 and FIBR 0527023 to H.M.S). V.C.
and H.M.S. are grateful for generous support from DARPA/IPTO BioComp
program, contract number MIPR 03-M296-01. H.M.S and V.C thank Dr Carsten
Peterson and Dr Michael Kastan for useful discussions.

\pagebreak

\section*{Supplementary Materials}

In this supplement we describe reaction schemes, parameter values,
and supplementary figures referred to in the main manuscript.

\subsection*{\atm~Switch Dynamics}
The reaction scheme, from which Eq 2, in the main text, are
derived is described by,
\medskip

\begin{center}
\begin{tabular}{ll} \hline\\
$\atm \rightarrow \atmp$ & $\frac{[\atm][\nbsp]}{k_{1} + [\atm]}+\alpha $ \\[6pt]
$ \atmp \rightarrow  \atm $& $\frac{[\atmp]}{k_{2} +[ \atmp]} \left(\frac{\alpha_s}{k_{0}+[\dam]}\right)$,\\[6pt]
$ \nbs \rightarrow \nbsp $& $\frac{[\nbs][\atmp]}{k_{4} + [\nbs]} +\beta $\\[6pt]
$\nbsp \rightarrow  \nbs $& $\frac{[\nbsp]}{k_{3} + [\nbsp]}$\\[6pt]
\\\hline
\end{tabular}

Reaction Scheme for the \atm~ Switch
\end{center}
\medskip
The parameter values for these equations are reported in Table.~1.
Each of \atm~and \nbs~are conserved cycles, i.e. \atm $+$\atmp~$ =
1$, and \nbs $+$\nbsp~$ = 1$
\begin{table}[h]
\begin{center}
\begin{tabular}{|c|c|c|c|c|c|c|c|}
\hline
$k_{1}$ & $k_{2}$ & $k_{3}$ &$k_{4}$  &  $k_0$ &$\alpha$ &$\beta$&$\alpha_s$\\
\hline
0.01& 1 & 0.1 &0.01& 0.1& 1E-4 &1E-4&0.3\\
\hline
\end{tabular}
\caption{Parameters used for the \atm~ switch.}
\end{center}
\end{table}

In the model for the \atm~switch, we assume a positive feedback,
by allowing \nbsp~to positively regulate the phosphorylation of
\atm~to \atmp. Here we show though simulation, the effect of
neglecting the positive feedback, on the dynamics of \atmp.
Although \atmp~levels rise rapidly with DNA damage, the decline
with subsequent DNA repair, is slow, and hence this would not
explain the switch-like behavior of \atmp. As is explained in the
main text, the rapid drop in \atmp~levels is required to turn {\bf
OFF} the \pft~oscillator.

\begin{figure}[h]
\begin{center}
\includegraphics[scale=0.5]{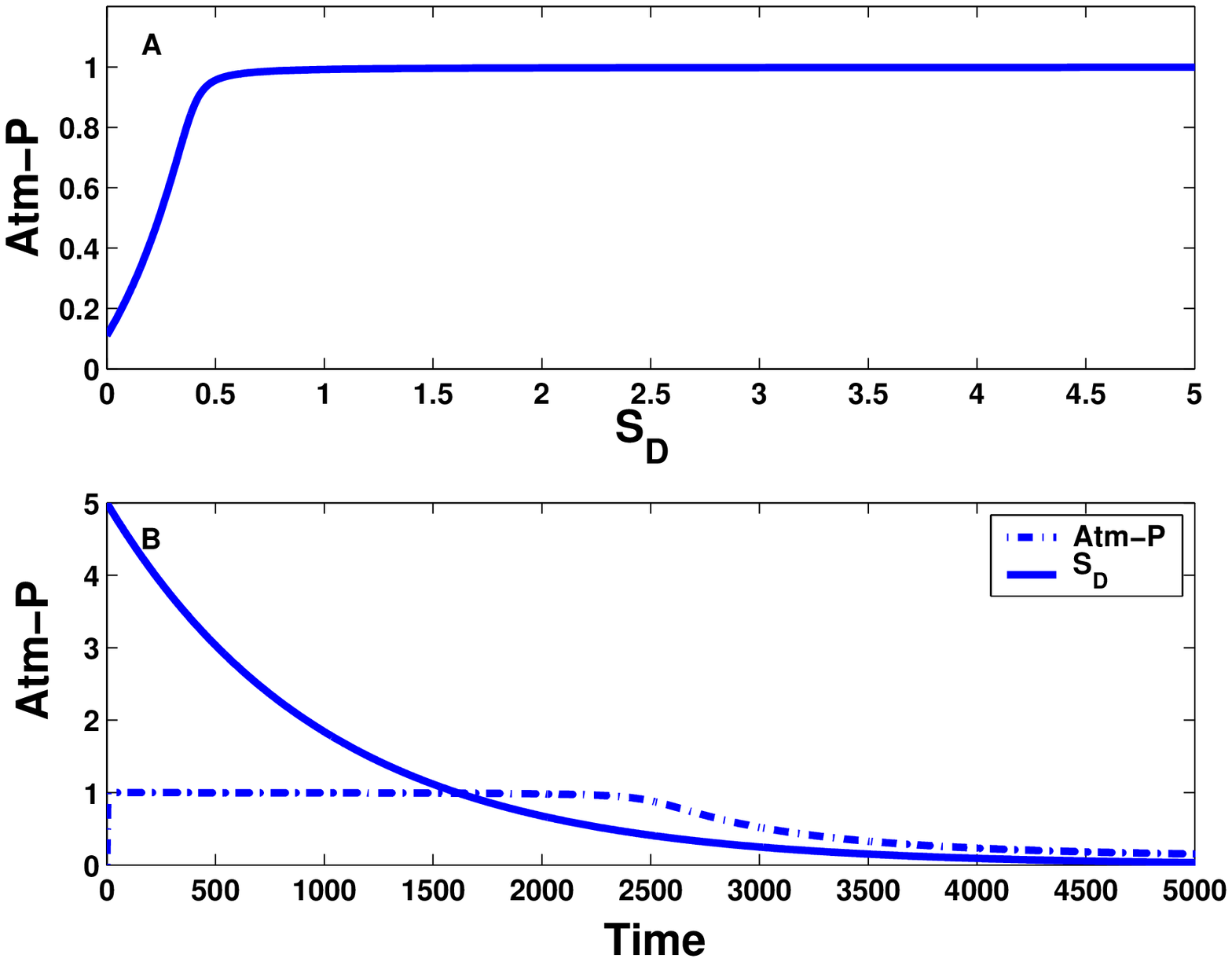}
\end{center}
\caption{In the upper plot is shown the steady state values of \atmp~as a
function of the DNA damage signal \dam, which shows ultrasensitive
behavior. The lower plot shows \atmp~as a function of time. \dam, whose
initial value is $\dam=5$, is modeled here, as an exponentially decreasing
function in time. Initially, \atmp~jumps almost immediately to a high level
($\simeq 1$), and then subsequently, as \dam~decreases it reduces to a low
value, but not rapidly enough, as is required by the integrated model.}
\end{figure}
If in the above reaction scheme, we leave out the positive
feedback, in the activation of \atmp~ by using $\atm \rightarrow
\atmp$ , with the rate law, $\frac{[\atm]}{k_{1p} + [\atm]}+\alpha
$, then a simulation of the switch model leads to the results
shown in Fig. 7. The dynamics of \atmp~does not rapidly decline as
the damage decreases. The latter effect is required to turn off
the \pft~oscillator.

In the main text in Fig.~6, we show plots of the steady state
values of \atmp, as a function of \dam, for two cases. case A:
Modified phosphatase activity, with $k_{0}=1$. case B: Increase in
the basal phosphorylation rates, $\alpha=0.01, \beta=0.01$. The
rest of parameters remain the same for both cases (as given in
Table 1).

\subsection*{\pft~Oscillator Dynamics}~~The equations for the \pft~oscillator model given in Eq. 3, in the
main text are obtained from the following reaction
scheme,~~\begin{center}~\begin{tabular}{ll} \hline\\
$source \rightarrow \pft$ & $c_0$ \\[6pt]
$\pft \rightarrow\pftp$  & $\atmp \frac{[\pft]}{(k_p+[\pft]}+con$ \\[6pt]
$\pftp \rightarrow \pft$  &  $P_h \frac{[\pftp]}{(k_{dep}+[\pftp])}$ \\[6pt]
$ \pftp \rightarrow sink $ &  $k_d [\pftp]$ \\[6pt]
$source \rightarrow \Mdm  $&  $P1 \frac{[\pftp]^4}{(k_m+[\pftp]^4)}$ \\[6pt]
$\Mdm + \arf \rightarrow  $\arf-\Mdm$  $ & $k_{4a} [\Mdm] [\arf]$ \\[6pt]
$ $\arf-\Mdm$ \rightarrow   sink + \arf $ &  $k_6 [$\arf-\Mdm$] $\\[6pt]
$\pft + \Mdm \rightarrow   $\pft-\Mdm $$ & $k_{1p} [\pft] [\Mdm]$ \\[6pt]
$$\pft-\Mdm$ \rightarrow \Mdm + sink $ &$ k_{c1} [$\pft-\Mdm$]$ \\[6pt]
 $\pft \rightarrow sink  $&  $k_{2p} [\pft] $\\[6pt]
 $source \rightarrow\beta\mbox{-catenin} $ & $ p_1 $\\[6pt]
$source \rightarrow \siah $  & $Ps \frac{[\pftp]^4}{(k_s+[\pftp]^4)}$ \\[6pt]
$ \siah + \beta\mbox{-catenin} \rightarrow \beta\mbox{-catenin-}\siah $ & $k_{c4} [\siah] [\beta\mbox{-catenin}]$ \\[6pt]
$\beta\mbox{-catenin-}\siah \rightarrow \siah + sink  $ & $k_{c3} [\beta\mbox{-catenin-}\siah] $\\[6pt]
 $\arf \rightarrow sink$ &$ k_7 [\arf]$ \\[6pt]
$source \rightarrow \arf $  & $Pb [\beta\mbox{-catenin}] $\\[6pt]
 $\siah \rightarrow sink $ & $ k_{si} [\siah] $\\[6pt]
 $\beta\mbox{-catenin} \rightarrow sink $  & $k_8 [\beta\mbox{-catenin}]$ \\[6pt]
$\pftp + \Mdm \rightarrow \pftp\Mdm $  &$ k_{p3} [\pftp] [\Mdm]$ \\[6pt]
$\pftp \Mdm \rightarrow \Mdm + sink  $ & $k_{c2} [\pftp] [\Mdm]$ \\[6pt]
$ \Mdm \rightarrow sink  $&  $k_5 [\Mdm]$ \\[6pt]
\\ \hline
\end{tabular}
\end{center}
where the parameter values are given in Table 2,
\begin{table}[h]
\begin{center}
\begin{tabular}{|c|c|c|c|c|c|c|c|c|c|c|c|}
\hline
$k_{p3}$ & $k_p$ & con & $ P_h$  &  $k_{dep}$  & $k_d$& $k_{1p}$  &  $k_{2p}$ &  $k_{c1}$ &  $k_{c2}$ &$P_1$&$k_m$\\
\hline
0.005&0.4  & 1E-4& 1E-6&1& 0.006 & 1&0.5 &1 &1 &0.4&1\\
\hline\hline
$k_5$& $k_{4a}$ & $k_s$& $k_6$ &  $p_0$  &  $k_7$ & $p_3$  &  $k_{c3}$& $k_{c4}$ & $p_1$ &$k_8$&$k_{si}$\\
\hline
1E-4& 15&1E+5&2&0.01&0.3&0.1&2&3 &1&0.2&0.02\\
\hline
\end{tabular}
\caption{Parameters used in the equations.}
\end{center}
\end{table},

and where the constitutive rate of production of \pft~is $c_0=0.8$. In Fig.
8A, we plot the bifurcation plot for the steady states of \pftp, as a
function of \atmp. There are two supercritical Hopf bifurcations, between
which the system exhibits oscillations. Note however, that the minimum
value of the \pftp~oscillations do not tend to a low value, as is
experimentally observed. The time series of the concentrations for \pft,
\Mdm~are displayed in Fig. 8B, for $\atmp~=1.0$.
\begin{figure}[hbtp]
  \vspace{9pt}
\centerline{\hbox{ \hspace{0.0in}
\includegraphics[scale=0.45]{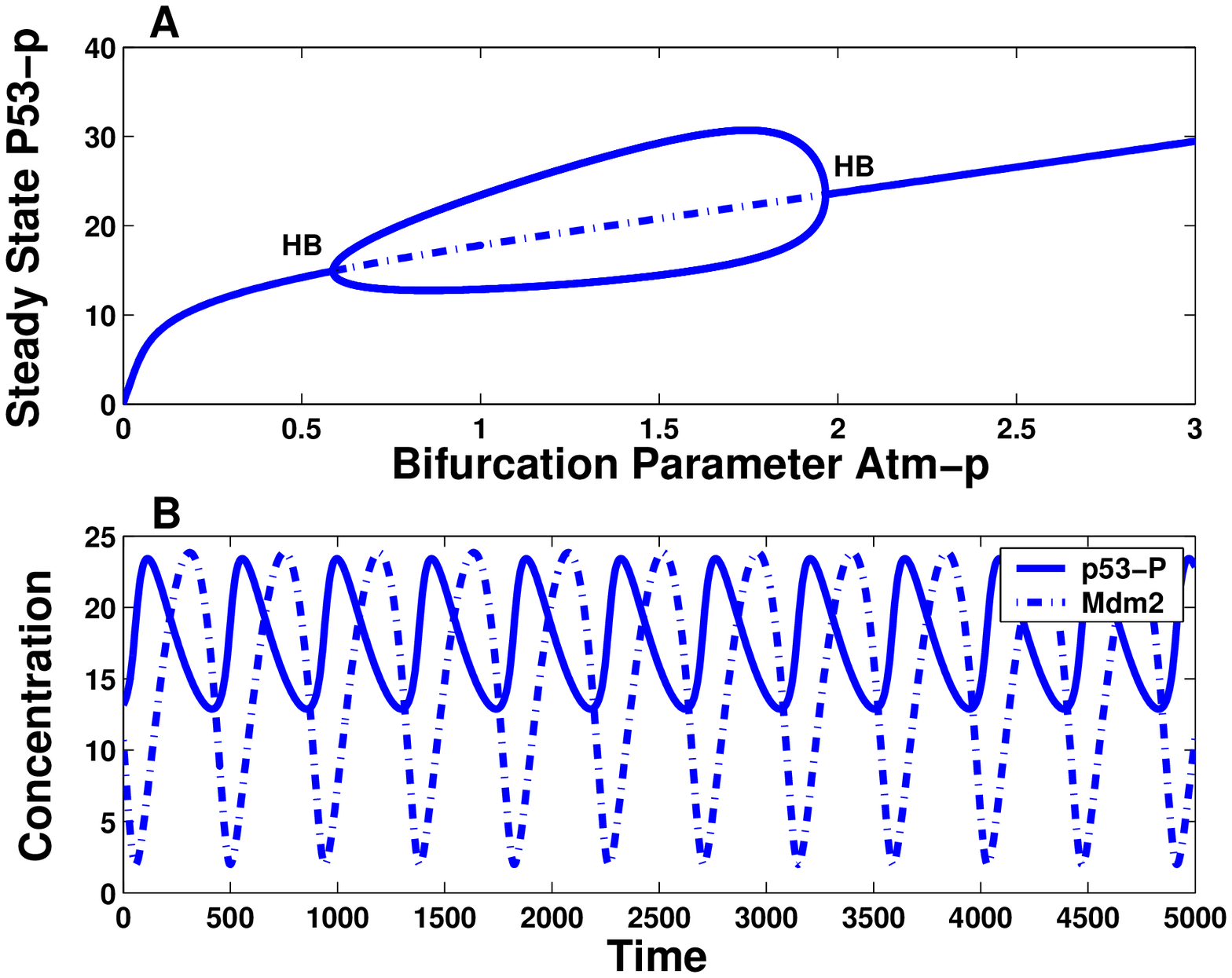}
    \hspace{0.2in}
    }
  }
  \vspace{9pt}
\caption{ Panel A shows the steady state value of \pftp~as a function of
\atmp~which shows two supercritical Hopf bifurcations (indicated by point
HB) that occurs at $\atmp \simeq 0.6, 2.0$. Within this region, the system
develops stable oscillations. The limits of the oscillations are marked by
the thick lines. The dashed line between the points marked HB indicate the
unstable branch. In panel B, we display
 the time series of \pftp~and \Mdm~, for a value of $\atmp=1.0$.}
\end{figure}
\subsection*{Sensitivity Analysis for module C}
In a sensitivity analysis of a network, the control coefficients
$C^{i}_{j}$ are typically computed,
\begin{equation}
C^{i}_{j}=\frac{p_j}{S_i} \frac{\partial S_i}{\partial p_j}
\end{equation}

where $S_i$, $p_j$ are the species, and the parameters respectively. Hence
$C^{i}_{j}$ expresses a fractional change in a species value, as one of the
parameters is changed. The coefficient could be positive, or negative,
depending on whether a small increase in a parameter tends to increase or
decrease a species level. The sensitivity is obtained by linearizing the
system about a steady state. In the case of the oscillator module, the
system oscillates about an unstable point. We fix the value of \atmp~$=1$,
for which the steady state can be computed. The computation is done using
the sensitivity features in JDesigner. In the table below, we compute the
control coefficients for a few important parameters for the oscillator
network (we display only those control coefficients that are either
reasonably large, or are important for regulation) , module C, which
describe the regulation that occurs.
\begin{table}[h]
\begin{center}
\begin{tabular}{|c|c|c|c|c|c|c|c|c|c|c|c|}
\hline
$C^{\pftp}_{c_0}$&$C^{\pftp}_{\atmp}$ &$C^{\pftp}_{P_1}$ &  $C^{\pftp}_{P_s}$  &  $C^{\pftp}_{P_b}$ \\
\hline
0.16&0.15  & -0.26&-0.24&0.26\\
\hline
\end{tabular}
\caption{Control coefficients for some of the important regulatory
steps in module C.}
\end{center}
\end{table}.

 For example we see that $C^{\pftp}_{c_0}$ is positive, which is
 intuitively obvious, since the parameter $c_0$ determines the
 production of \pft. The control coefficients $C^{\pftp}_{P_1},
 C^{\pftp}_{P_s}$, are approximately of same order, and of the same
 sign. An increase in $P_1, P_s$ represents an increase in the
 transcription of MDM2, SIAH genes, both of which lead to a reduction in
 \pftp. Hence a reduction in either transcription rate, can be
 compensated by an increase in the other. $C^{\pftp}_{P_b}$ is
 positive due to the following: an increase in $P_b$, leads to an increase in \arf,
 which sequesters more of \Mdm~and hence reduces the degradation
 of \pft~which therefore increases the amount of phosphorylated
 \pft.

\end{document}